\documentclass[reprint,secnumarabic,amssymb, nofootinbib, aps, prd,10pt]{revtex4-1}
\usepackage{graphicx}				
\usepackage{amssymb}
\usepackage{amsmath}
\usepackage{amstext}
\usepackage{setspace}
\usepackage{bbold}
\usepackage{ulem}
\usepackage{slashed}

\newcommand{\be}{\begin{eqnarray}}
\newcommand{\ee}{\end{eqnarray}}

\newcommand{\ep}{\epsilon}

\newcommand{\ttb}{t\bar{t}}

\begin{document} 
\title{Rare exclusive hadronic $W$ decays in a $\ttb$ environment}
    \author{Michelangelo Mangano and Tom Melia}
    \affiliation{CERN TH group, PH Department, CH-1211\\
    Geneva 23, Switzerland.}
    
    \begin{abstract}
The large cross section for $\ttb$ production at the LHC and at any future hadron collider
provides a high-statistics and relatively clean environment for a study of $W$ boson properties: after tagging on
a leptonic decay of one of the $W$s and the two $b$ jets, an additional $W$ still remains in the event.
We study the prospect of making the first exclusive hadronic decay of a fundamental boson of the standard model, using  
the decay modes $W\to \pi \gamma$ and $W \to \pi \pi \pi$, and other related decays. 
By using strong isolation criteria, which we impose by searching for jets with a single particle constituent, 
 we show that the three particle hadronic $W$ decays have potential to be measured at the LHC.
  The possibility of measuring an involved spectrum of decay products could considerably expand our
  knowledge of how the $W$ decays, and experimental techniques acquired in making these measurements would
  be useful for application to future measurements of exclusive hadronic Higgs boson decays.
    \end{abstract}
        
    \maketitle

Experimental measurements of the decay modes of the $W$ boson 
are summarised in the Particle Data Group (PDG) Review \cite{Beringer:1900zz}. They consist of: the three 
leptonic decays to $e,\mu,$ and $\tau$
plus a neutrino (with $\sim$1\% precision), and the total decay rate to leptons; inclusive hadronic decay ($\sim$0.5\% precision), which is split further
 into inclusive hadronic decays to  $cX$, and $c\bar{s}$; and invisible decay (consistent with zero at a level of $\sim$3\%). In addition to this, two 95\% confidence level 
 upper limits are set on the decay rate, $\Gamma_i$, for $W^+\to \pi^+\gamma$ ($\Gamma_{\pi\gamma}/ \Gamma_{\text{tot}} < 8 \times 10^{-5}$)\footnote{The CDF experiment improves on the limit quoted
 in the current PDG by an order of magnitude to $<6.4\times 10^{-6}$ \cite{Aaltonen:2011nn}.} and 
 $W^+\to D_s^+\gamma$  ($\Gamma_{D_s\gamma}/ \Gamma_{\text{tot}} < 1.3 \times 10^{-3}$), making up  a total of 10 entries in the table.

This is to be compared with the PDG table for the $Z$ boson, which reports over 50 different searches and measurements
of the decay modes of this particle, including semi-exclusive hadronic final states (e.g. $Z\to J/\psi \,X$, $D^\pm\, X$, $B^0_s\, X$, etc.), 
as well as upper limits on fully exclusive hadronic decays (e.g. $Z\to\pi^0\gamma $, $\pi^\pm W^\mp$) 
and on lepton flavour, lepton number
and baryon number violating decays (e.g. $Z\to\mu e$, $ep$). The leptonic decays and the
total inclusive hadronic decay of the $Z$
have been measured with a precision over an order of magnitude better than those of the $W$, i.e. at the per mille level. The difference
in the PDG tables
reflects the fact that LEP, being an electron-positron collider, could singly produce of order $10^7$ $Z$ bosons in an experimentally
clean environment. Although $\sim10^{11}$ $W$ bosons will be produced at the high luminosity (HL) Large Hadron 
Collider (LHC), and orders of magnitude more at proposed future hadron colliders,
the huge QCD background to generic $W$-production final states, and
the  trigger challenges, render many precision studies of $W$
decays implausible at these machines. 
Proposed future electron-positron colliders
will pair produce $W$ bosons in a clean
environment, but, 
even in the case of circular accelerators such as
TLEP~\cite{Gomez-Ceballos:2013zzn} or CEPC~\cite{cepc}, 
they at best promise samples of $\sim10^8$ events.
Can the hadron collider experimental barrier
be overcome and the huge statistics be exploited?
  \begin{figure}
\begin{center}
\includegraphics[width=8.2cm]{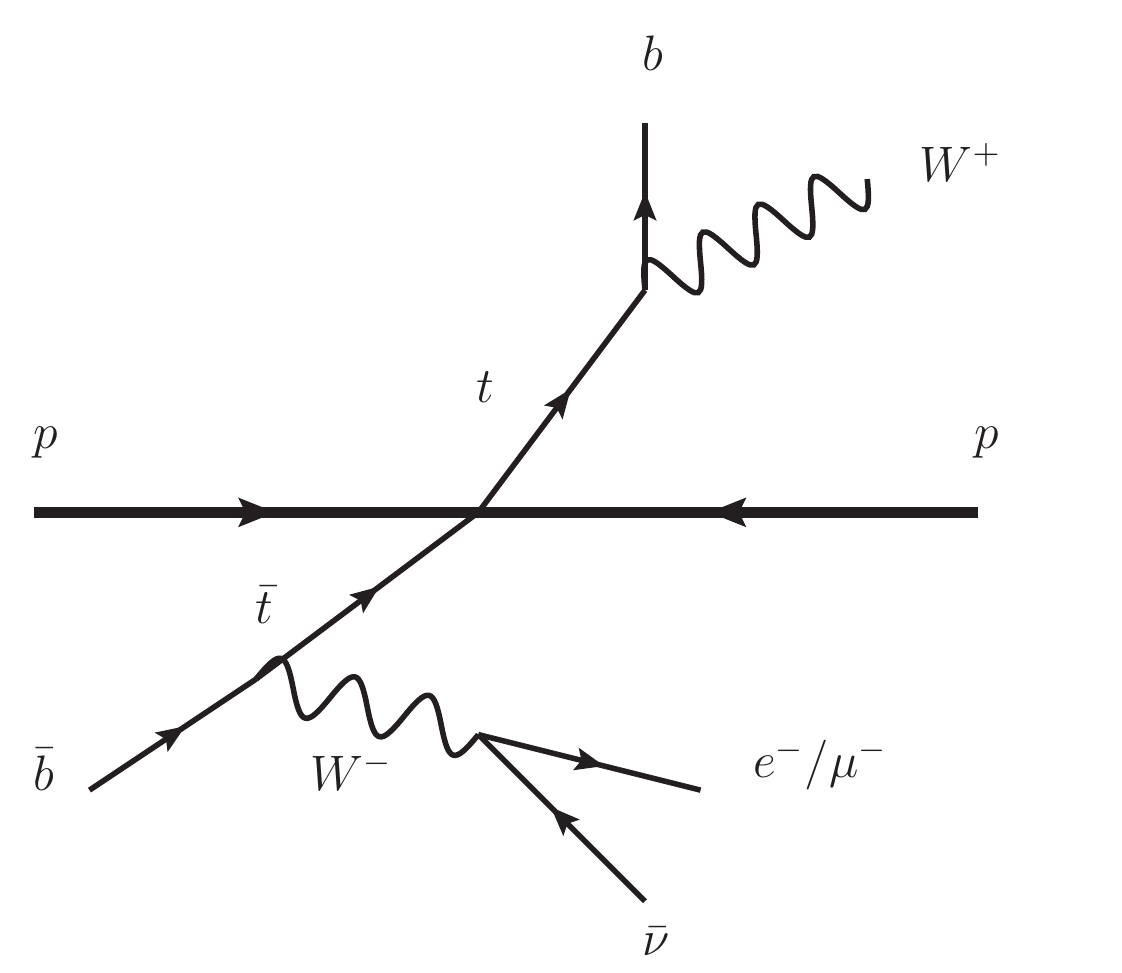}
\end{center}
\caption{The $\ttb$ environment in which the $b$-jets are tagged on, and the leptonic decay
of one of the $W$s is required.}
\label{fig:ttbar}
\end{figure}

One of the main ideas in this note is to highlight that the enormous
$\ttb$ production cross section at hadron colliders operating at LHC
energies and above is a  promising environment in which to make
precision measurements of the $W$ boson, given the manageable QCD
background and given the trigger opportunities.  Top quarks decay
dominantly into a $b$ quark and a $W$, and by tagging on the leptonic
decay of one of the $W$ bosons in the event, as well the $b$-jets, a
situation is created where inclusive decays of the leftover $W$ boson
in the event can be studied in an rather unbiased way -- see
Fig.~\ref{fig:ttbar}.  There will be $O(10^{9})$ $W$ bosons
potentially triggerable
in this way at the end of the HL-LHC run, and $O(10^{11})$ $W$s at a
100 TeV collider taking 10ab$^{-1}$ of data, i.e. orders of magnitude
more than $Z$ bosons at LEP or $W$ bosons at a future circular
$e^+e^-$ collider.  This opens a possible door to a high-statistics program
of $W$ boson studies, including searches for
rare or forbidden decays (e.g. lepton flavour and lepton number
violating decays~\cite{BarShalom:2006bv}), 
and improved measurements of leptonic and hadronic
branching ratios.

In this note we focus on fully exclusive hadronic decays of the $W$,
which are experimentally very difficult to study at a hadron
collider. For this we use a technique that utilises what can be seen
as an extreme form of jet substructure and which we refer to as single
particle jet isolation -- requiring jets which have as constituents a
single particle (with a more loose definition when the particle is a
photon).  This method relies on the fact that a well isolated single
hadron or photon is a rare outcome of generic QCD evolution. This
approach is clearly analogous to what is done experimentally to
identify hadronic decays of tau leptons.

Of the three massive fundamental bosons of the
standard model, not a single exclusive hadronic decay mode has ever
been measured. Low-multiplicity decays can only arise from a
perturbative evolution of the final state with radiation of few (or no)
gluons, with a probability that is greatly suppressed by
Sudakov effects in the form of powers of $\Lambda_{QCD}/m_W$.
The observation of such decays would therefore probe
strong-interactions in a very interesting dynamical domain, at the
borderline of perturbative and non-perturbative physics. 
Furthermore, a number of recent
papers~\cite{Isidori:2013cla, Bodwin:2013gca,
  Kagan:2014ila,Gao:2014xlv,Bhattacharya:2014rra,Keung:1983ac} have
addressed the idea of using exclusive hadronic decays of the Higgs
boson $h\to VM$, where $V=W,Z,\gamma$ and $M$ is a meson, as a test of
both the on- and off-diagonal couplings of $h$ to quarks. Such
measurements are very challenging at the LHC, and observing exclusive
hadronic decays of the $W$ would provide a proof of principle that
this type of final state is accessible at a hadron collider. As
pointed out in \cite{Kagan:2014ila}, future electron-positron
colliders do not have the required statistics for observing these
decays.

We present a Monte Carlo (MC) study, performed at particle level, using single particle jets to
overcome the overwhelming hadronic activity at a hadron collider.
We focus on the phenomenologically simplest  two and three particle exclusive decays to mesons that are `stable' as far
 as the LHC detectors are concerned, 
 \be
 &W^+&\to \pi^+ \gamma \nonumber \\
 &W^+&\to \pi^+\pi^+\pi^- \nonumber
\ee
 and we show that requiring single particle jets provides  an excellent
  handle for separating signal from background. We highlight that three-particle decays in particular
  are  candidates for an LHC measurement. Related decays with pions substituted 
  by other charged particles, such as $K^+$, $D^+_s$ etc.
  are also discussed, as well as the prospect of mass measurement in these fully visible decay
  modes. 
   
We  proceed as follows: in Sec.~\ref{sec:one} we discuss theoretical issues surrounding exclusive hadronic decays of 
weak bosons. In Sec.~\ref{sec:two} we present a MC particle level study which uses  the  technique
of single particle jet isolation to measure these decays at hadron colliders, and comment on further
experimental handles that can be used to increase sensitivity to these decays in the $\ttb$ environment. 
In Sec.~\ref{sec:disc} we present our conclusions and outlook, including implications of our results found
for exclusive Higgs boson decay.

\section{Rare Exclusive Hadronic Decays of the $W$ Boson}
\label{sec:one}
  \begin{figure}
\begin{center}
\includegraphics[width=9cm]{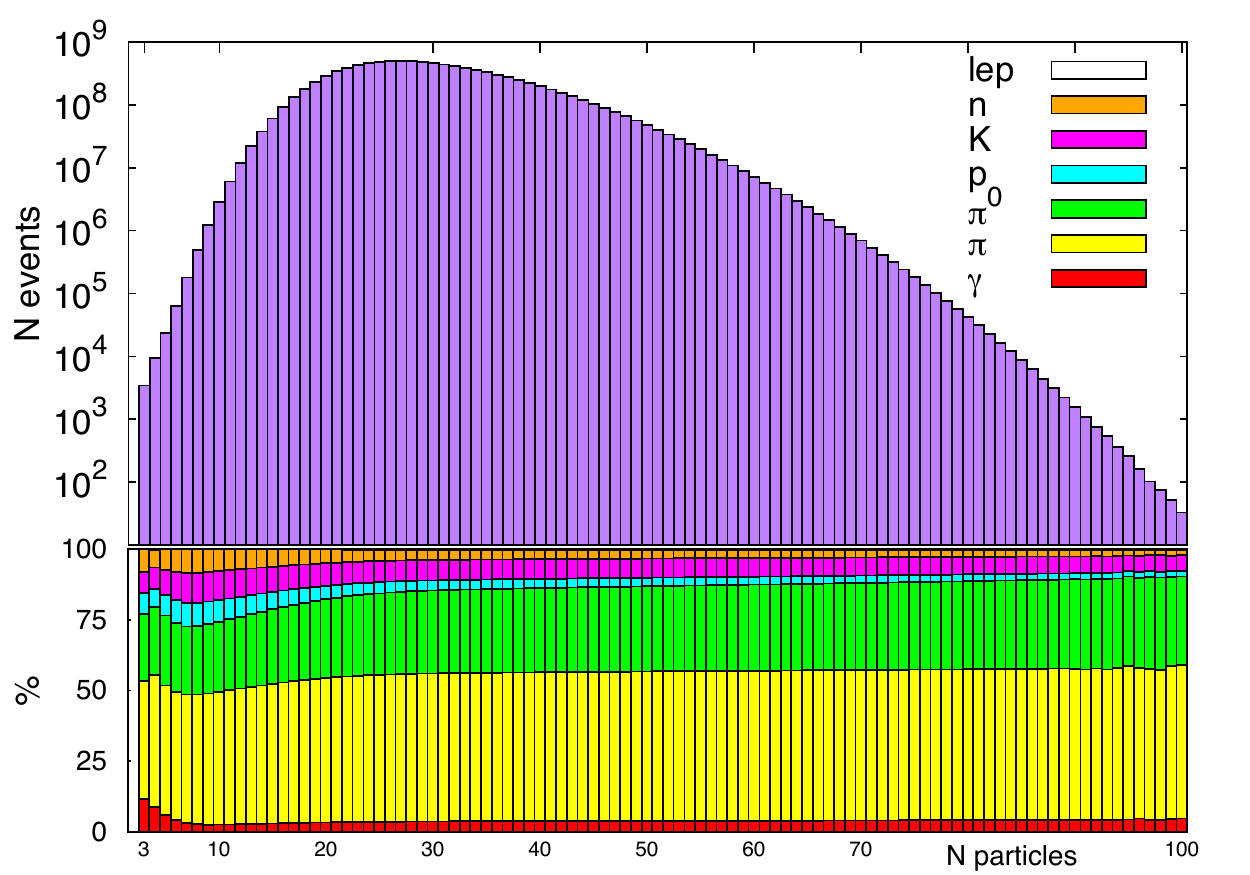}
\end{center}
\caption{The distribution of the
 number of final state particles in $10^{10}$ showered and hadronized $W\to u\bar{d}$ decays  in PYTHIA 8. Top panel
 gives total number of events. The legend corresponds to the bottom panel where the average fraction of types of particle
 in the final state are given (in descending order): leptons (not visible on plot),  neutrons, kaons (both charged and neutral), protons, neutral pions, charged pions, and photons.}
\label{fig:pith}
\end{figure}

The main reason that no exclusive hadronic final state of the weak
bosons, $W$ or $Z$, has ever been observed is because the majority of
the decays are into $\sim30$ particle final states (as seen by the
detector), composed of charged and neutral pions and kaons, protons,
neutrons, photons, and leptons. To get a feel for the distribution of
final states, we show in Fig.~\ref{fig:pith} the result of decaying,
showering and hadronizing
$10^{10}$ $W^+\to u\bar{d}$ with PYTHIA 8 \cite{Sjostrand:2006za,
  Sjostrand:2007gs}, letting any
resonances decay to particles seen in the detector (except for neutral
pions, which we keep undecayed -- these will decay
$\pi^0\to\gamma\gamma$) and counting the number of each type of
particle produced.  Clearly, a measurement of branching fractions to
any of the given exclusive, high-multiplicity final states that
dominate the decay is implausible, especially as many of the decay products are
neutral and hard to identify. However, PYTHIA does find that a number of
three particle final states are possible and can be found in the MC
final state:
\be 
&W^+& \to \pi^+ \pi^+ \pi^-
\nonumber\\ 
&W^+& \to \pi^+ p \, \bar{p} \nonumber \\ 
&W^+ & \to \pi^+ K^+ K^- \nonumber\\ 
&W^+ & \to \pi^+ n^0\, \bar{n}^0 \nonumber\\ 
&W^+ & \to \pi^+ \pi^0 \gamma \nonumber\\ 
&\ldots& \nonumber 
\ee 
some of
which contain only charged hadrons. The rate at which these final
states occur is one every $10^6-10^7$ events, but the PYTHIA models
are constrained by charged multiplicity data from LEP measurements of
$Z\to$ hadrons which have large errors in the extremities of the
multiplicity distributions (for the most recent LHC PYTHIA tune and a
detailed discussion, see \cite{Skands:2014pea}, and references
within).  We expect from the perturbative QCD picture that the decay
$W\to \pi\pi\pi$ would have a rate of the same order of magnitude as
$Z\to \pi^0 \pi\pi$. We are not aware of direct searches for such a
final state, and cannot infer, from the available information on the
study of hadronic final states at LEP, what the constraints on its
branching fraction could be. Given $\sim 10^7$ $Z$ bosons were produced at LEP,
however, no constraint below $O(10^{-6})$ can likely be obtained.
The value obtained above with PYTHIA is therefore consistent with
LEP. It is possible that the decay rate could be as large as $\sim 10^{-5}$
and still be within experimental bounds.

\begin{figure}
\begin{center}
\includegraphics[width=9.8cm]{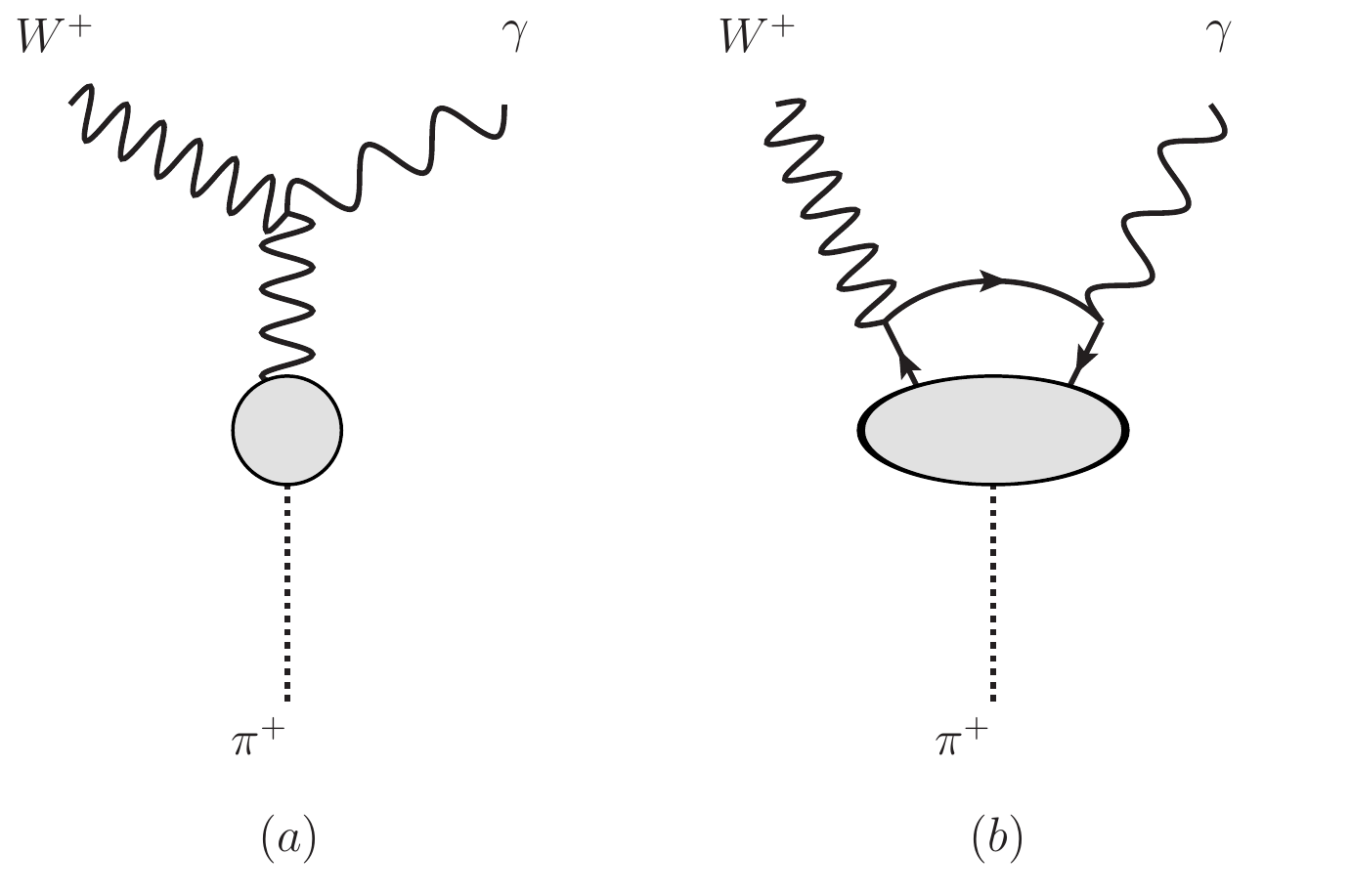}
\end{center}
\caption{Contributions to the decay $W^+ \to \pi^+ \gamma$.}
\label{fig:amps}
\end{figure}

The two particle decay mode $W^+\to\pi^+\gamma$ was not found by
PYTHIA in the above exercise of showering and hadronizing $W^+\to
u\bar{d}$.  This region is very extreme as it requires the $u$ and
$\bar{d}$ quark to be recoiling against a photon, with a very small invariant
mass.  What is more, a contribution to the decay where the
photon couples directly to the $W$ boson, shown in
Fig.~\ref{fig:amps}~(a), is not taken into account.
The rate can be estimated as follows. Contribution (a) can be related directly to the pion decay constant, $f_\pi=93\,$MeV, 
via an evaluation of the current
\be
 \langle \pi^+(p) | J_W^\rho(0) | 0 \rangle = \frac{f_\pi}{\sqrt{2}}\,\, p^\rho\,,
\label{eq:one}
\ee
where $p$ is the momentum of the pion state $|\pi^+(p)\rangle$, and $J_W^\rho = \bar{d}\gamma^\rho P_L u$ is the weak current, with $P_L=\frac{1}{2}(1-\gamma_5)$. Contributions of the type shown in Fig.~\ref{fig:amps}~(b) involve a calculation of
\be
\int{d^4 x e^{i k\cdot x} \langle \pi^+(p) | T[ J_W^\lambda(0) J_\gamma^\mu(x)] | 0 \rangle}
\label{eq:two}
\ee
where $k$ is the photon momentum, and $J_\gamma^\mu = \sum_{i=u,d}Q_i\bar{q}_i\gamma^\mu q_i$ is 
the electromagnetic current, with $Q_i$ the charge
of the quark $q_i$.
To evaluate these contributions we adapt 
Manohar's calculation of the decay width $Z\to W^\pm\pi^\mp$,
and subsequent estimate of the decay $Z\to \pi^0 \gamma$ \cite{Manohar:1990hu}, which uses
 an operator product expansion (OPE) at leading order in 
the strong coupling constant $\alpha_S$, retaining only the leading terms in a tower of twist two operators. 
We review this calculation in the appendix. We obtain an order of magnitude estimate for 
$\Gamma_{\pi\gamma}/\Gamma_{\text{tot}}\sim 10^{-9}$, although the expansion is not convergent and
 will be modified by important higher order corrections. 
 This result can be compared
to previous results in the literature for this decay. 
A calculation by Arnellos, Marciano and Parsa (AMP) \cite{Arnellos:1981gy} also yields
 a value for $\Gamma_{\pi\gamma}/\Gamma_{\text{tot}}$ 
 around $10^{-9}$, assuming the Brodsky-Lepage (BL)
asymptotic formula \cite{Lepage:1979zb} for the off-shell photon-photon-pion vertex, $\gamma^\star \gamma \pi$, for both the vector and axial
form factors\footnote{This asymptotic limit for $\gamma^\star \gamma \pi$ is defined when the 
mass of the off-shell photon $Q^2\to\infty$. However, it too receives important higher order terms
when $Q^2$ is very large but finite \cite{Manohar:1990hu}. This form factor is  relevant for the off-shell 
$\gamma^\star\gamma \to \pi$ production anomaly
at Belle \cite{Uehara:2012ag} and BaBar \cite{Aubert:2009mc}, see e.g.
\cite{Stefanis:2012yw,Guo:2011tq,Polyakov:2009je,Klopot:2012hd,McKeen:2011aa,Roig:2014uja,Pham:2011zi,Masjuan:2012wy,Brodsky:2011yv}.}.
A one-loop calculation by Keum and Pham \cite{Keum:1993eb} gives a
prediction of
$\Gamma_{\pi\gamma}/\Gamma_{\text{tot}}\sim10^{-8}-10^{-6}$.  This
calculation follows closely the very similar calculation of the famous
$\pi^0 \to \gamma \gamma$ anomaly \cite{Adler:1969gk,Bell:1969ts} and
is obtained via a use of the Goldberger-Treiman relation
\cite{Goldberger:1958vp} for the quark-quark-pion vertex, yielding
$\Gamma \propto m_q^4$, where $m_q$ is the quark mass. The upper value
of $10^{-6}$ follows from using current quark masses $m_q\sim300\,$MeV
in the loop. The Golberger-Treiman relation is not valid if the
momentum running in the quark loop is truly at the scale of the $W$
mass, i.e. much greater than $4\pi f_\pi$. Because of this, we expect that the
upper estimate of $10^{-6}$ is too large (furthermore, we expect the
$\pi\gamma$ branching ratio to be considerably smaller -- at least by
a factor of $\alpha_{EM}$ -- than that the $\pi\pi\pi$ branching ratio,
which, from LEP, is experimentally unlikely to exceed
$\sim10^{-5}$).

Is it possible to observe final states with branching ratios as small
as those described above?  We now turn to the challenge of observing
them in high energy proton-proton collisions, where pions and photons
are produced in huge numbers, and study the use of single particle jet
isolation in the $\ttb$ environment.

\section{ Single particle jet isolation}
\label{sec:two}

 We perform a MC study to estimate the reach of the HL-LHC and future hadron colliders
 in observing such two and three particle exclusive
 hadronic decays, 
 with two separate analyses to search for $W\to \pi\gamma$ and $W\to \pi\pi\pi$ as explicit examples. 
 The numbers presented in this section are for 14 TeV $pp$ collisions, and we discuss the scale-up to 100 TeV
 in the following section. 
 We
  generate $\ttb$ events where one of the $W$s decays as $W^-\to e^- \bar{\nu}_e$,
 using MadGraph 5 \cite{Alwall:2014hca} at tree-level. The two separate signal samples are generated by then 
 forcing the $W^+$ to decay to $\pi^+\gamma$ or $\pi^+\pi^+\pi^-$ isotropically (neglecting spin effects). 
 A background sample is generated by allowing
 the $W^+$ to decay generically hadronically -- we refer to this as the `$W$-had' background\footnote{A check 
 was put in place to disregard any of these events where the $W^+$ decays into 
a two or three particle state after hadronisation.}. We  separately  generate the background where the $W^+$ decays
into a tau lepton, referred to as the `$W$-tau' background, where the tau can then decay hadronically. 
Since a tau decays to one or three charged pions and kaons
$\sim12\%$ and $\sim15\%$ of the time respectively, this could be an 
important background.
We also considered  QCD $W^-b\bar{b}$ production, with $W^-\to e^- \bar{\nu}_e$, which is an 
irreducible background to the tagging procedure
for the $\ttb$ environment; this background is subdominant to the ones above (as well as displaying a very
different event topology to the $\ttb$ one, which could be utilised to suppress it further) and we discuss it no further here. 
In all samples, the electron from the $W$ decay is required
to be separated from the $b$ partons by $\Delta R_{eb}>0.3$.
These samples are then showered and 
hadronized with PYTHIA 8, without the addition of pileup.
 \begin{figure}
\begin{center}
\includegraphics[width=9cm]{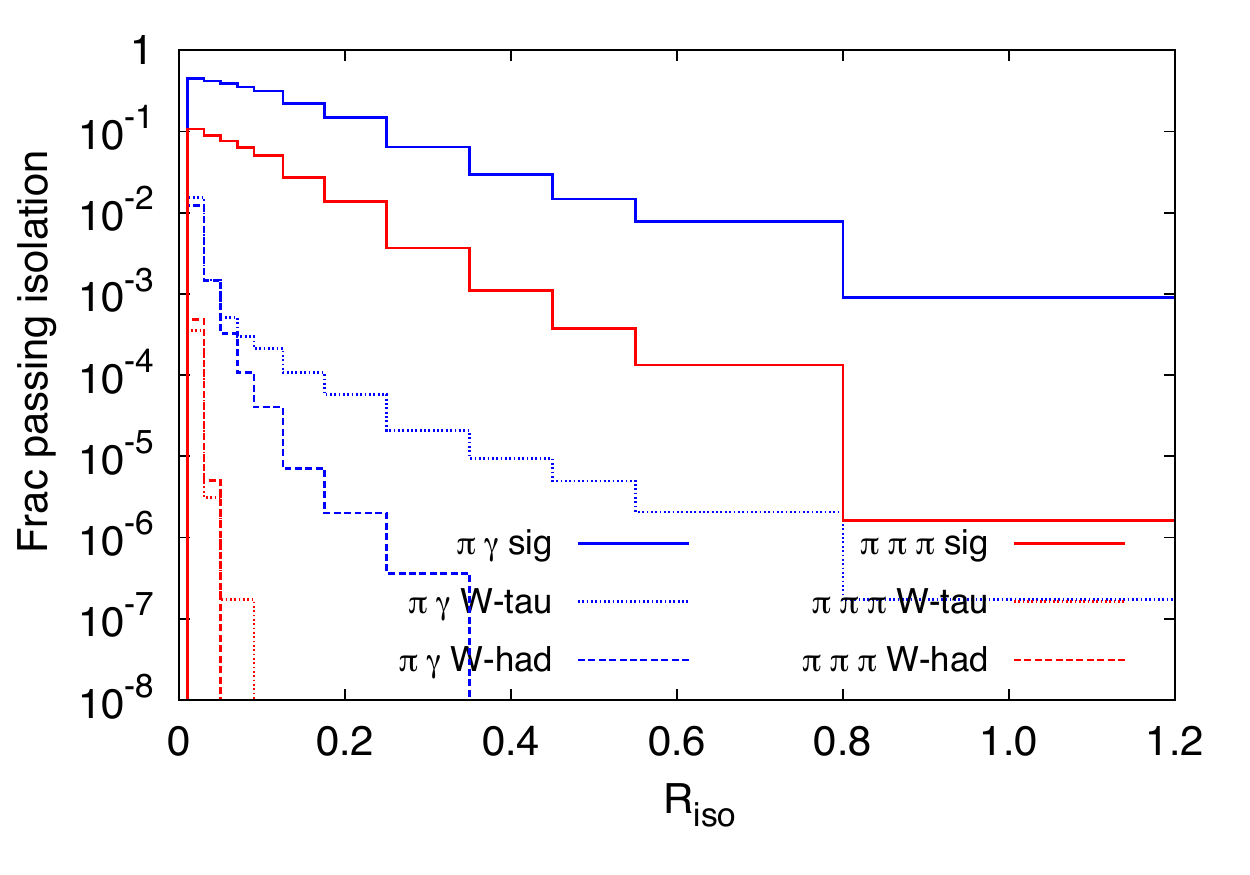}
\end{center}
\caption{Fraction of signal and backgrounds passing the single particle jet isolation
cuts as a function of the parameter $R_{iso}$.}
\label{fig:riso}
\end{figure}

The analysis involves the following steps, which we describe in detail below:
\begin{itemize}
\item Select $\ttb$ events by requiring an electron, missing energy, and two $b$-jets constructed with cone size $R=0.4$.
\item Remove all particles associated with the $b$-jets from the analysis.
\item Re-cluster the remaining particles using a cone size $R=R_{iso}$ and require a single particle jet (defined below) for 
each final state hadron of the decay.
\end{itemize}
For the triggering cuts, designed to select the $\ttb$ event, jets are reconstructed with FastJet \cite{Cacciari:2011ma}, 
using the anti-kT algorithm \cite{Cacciari:2008gp} with $R=0.4$,
and requiring the transverse momentum of the jet $p_T^j>25\,$GeV and
rapidity $|\eta_j|<2.5$. The constituents of these jets are analysed and a
 jet is considered $b$-tagged if any of its constituents is a $b$-hadron. We require two $b$-tagged jets.
The electron is required to have $p_T^e>20\,$GeV and $|\eta_e|<2.5$, and the missing transverse momentum $p_T^{\text{miss}}>30\,$GeV (constructed as the
modulus of the vector sum of  the transverse momentum of all final state particles with $|\eta|<3.6$, except for neutrinos).  
We choose not to impose a transverse mass cut on the top quark. Such a cut would suppress
any QCD background, but the background we consider here is found to be
suppressed enough by the following selection criteria, so it is
advantageous in this analysis to keep as much signal as possible.

If the event passes these triggering cuts, single particle isolation
cuts are then implemented to separate signal from background.
Firstly, all of the particles associated with the two $b$-tagged jets
are removed from the event.  The event is then resent to FastJet with
a different $R$ parameter, $R=R_{\text{iso}}$, which we vary in the
following, and we again construct jets with $p_T^j>25\,$GeV,
$|\eta_j|<2.5$. We call a jet a ``single pion jet'' if it is
composed of exactly one charged pion.  Similarly a ``single
$\gamma$''-jet is defined when all of the constituents of the jet are
photons. This definition is loose in the sense that no differentiation
is made between MC jets consisting of a single photon and those
containing two or more photons (for example coming from the decay of a
$\pi^0$). We assume for now charge and particle identification used in
the definition of the single pion jet. For the $\pi\gamma$ analysis,
events pass these selection cuts if at least one $\gamma$-jet and one
charged single pion jet are found. For the $\pi\pi\pi$
analysis, we require at least three  charged single pion jets. 
If more single particle jets are
found, the ones with the hardest transverse momentum are selected
(this happens a negligible amount of the time).

\begin{figure}
\begin{center}
\includegraphics[width=9cm]{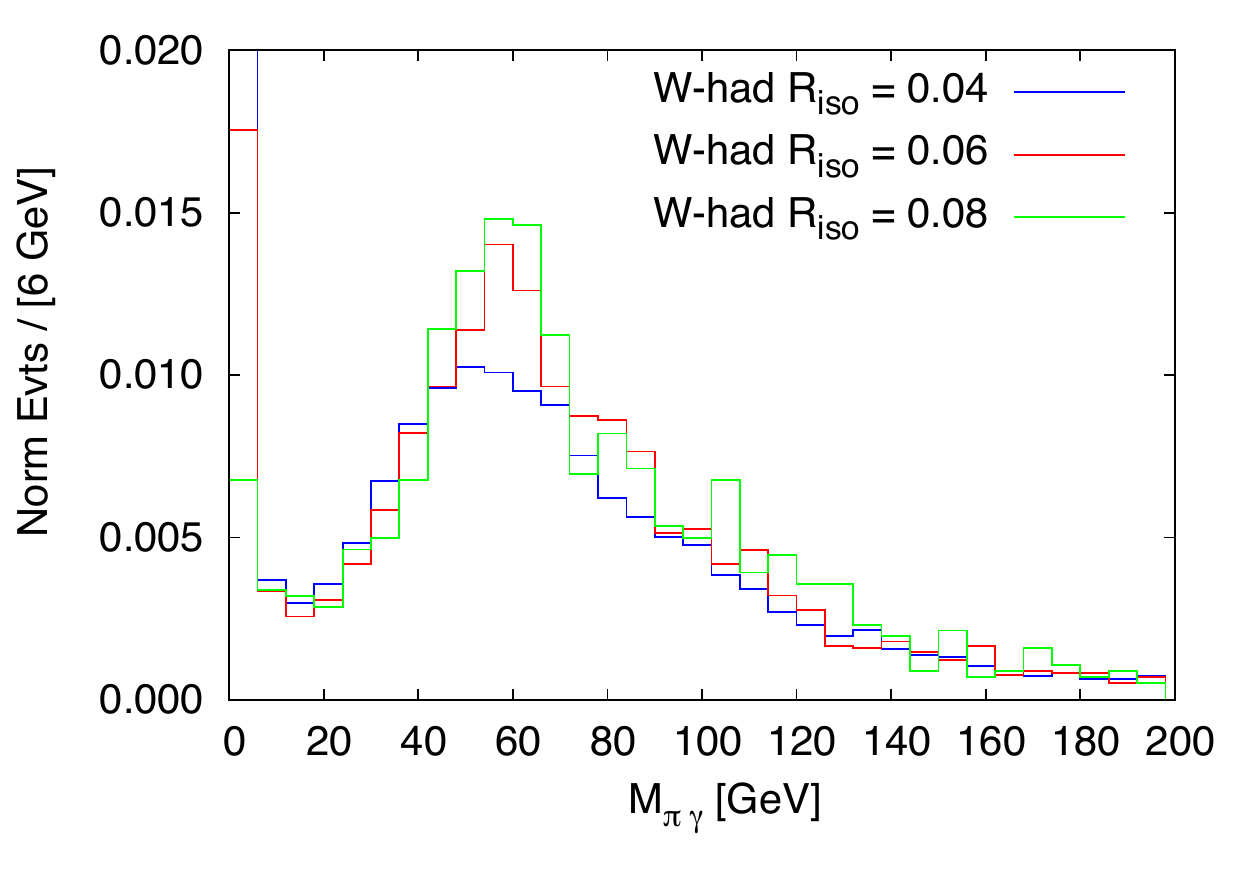}
\includegraphics[width=9cm]{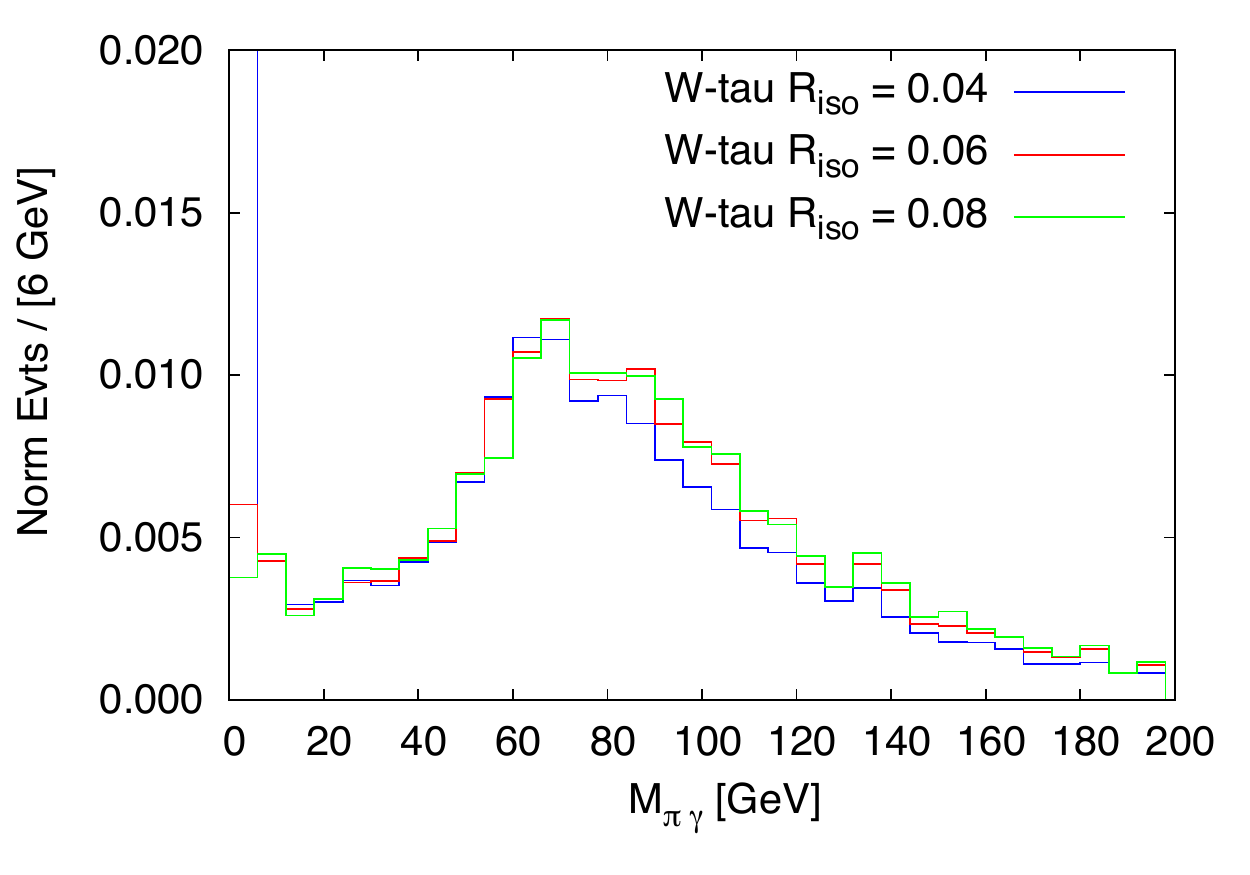}
\end{center}
\caption{Top: distribution of $M_{\pi\gamma}$ for the generically hadronically decaying $W$ background
in the $W\to \pi \gamma$ analysis, plotted
for different values of $R_{iso}$. 
Bottom: distribution of $M_{\pi\gamma}$ for the $W$ decaying to tau lepton background
in the $W\to \pi \gamma$ analysis, plotted
for different values of $R_{iso}$.  }
\label{fig:main}
\end{figure}

\begin{figure}
\begin{center}
\includegraphics[width=9cm]{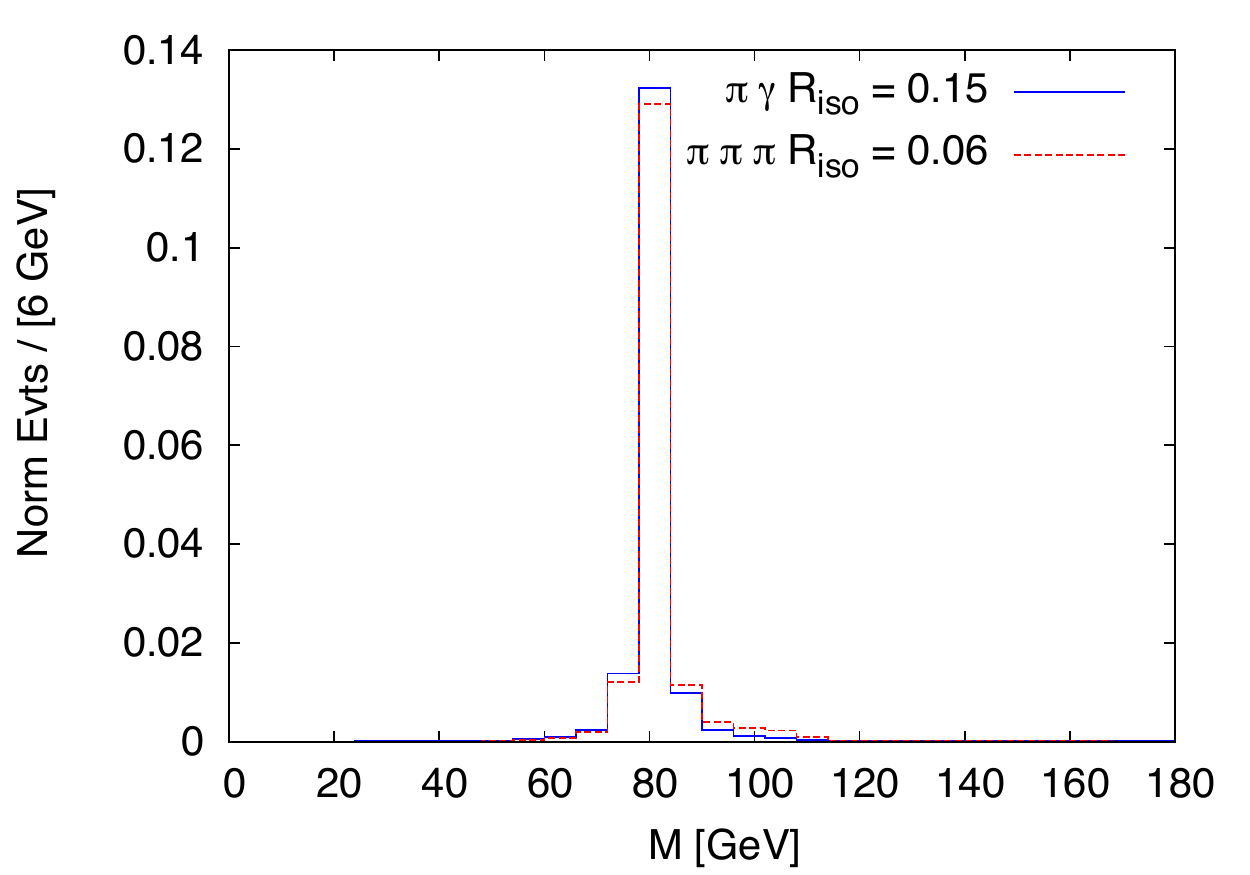}
\end{center}
\caption{Signal distribution of the reconstructed $W$ mass for the decays $W\to \pi\gamma$ and $W\to \pi\pi\pi$.}
\label{fig:sigvar}
\end{figure}

The fraction of signal, $W$-had background, and $W$-tau
background passing these isolation cuts 
as a function of the $R_{\text{iso}}$ parameter are shown in Fig.~\ref{fig:riso}. For both analyses, the $W$-had 
background falls off considerably faster than the signal as $R_{\text{iso}}$ is increased. Both backgrounds
fall so quickly in the $\pi\pi\pi$ case that 
there is essentially no background in this MC study for $R_{iso}\ge0.06$. 
The $W$-tau background does not fall off for large values of $R_{iso}$ in the $\pi\gamma$ analysis. This is caused by events
where the tau decays $\tau^+ \to \pi^+ \bar{\nu}_\tau$ ($\sim 10\%$ of the time) to create the single isolated pion, and where there
is a hard, well-separated photon radiated from elsewhere in the event (top, $b$-quark, initial state, electron). Since the event
is relatively empty (with three neutrinos), the cost of this is $\sim \alpha_{EM}$, and so this background
tracks the signal, being below it by a factor of $\sim10^{-1}\times 10^{-3}$. It is reducible to the extent to which the pion can be
identified as coming from a tau decay, using displaced vertex tagging,
although given that the tau will be well boosted in the laboratory frame we do not expect this to provide 
any significant experimental improvement\footnote{The analogue of this background does not affect
the $\pi\pi\pi$ analysis, since the boost of the $\tau$ means the decay $\tau\to\pi\pi\pi\nu$ is pencil-like.}. The
$\pi\pi\pi$ signal is seen to be less efficient than the $\pi\gamma$, simply because in this case 
three particles have to pass both transverse 
momentum and isolation cuts.
We plot the shape of the background 
$M_{\pi\gamma}$ distributions
  in the $\pi\gamma$
analysis in Fig.~\ref{fig:main}. The peak of the distributions is driven by the $p_T$ cut on the jets, and
lowering this cut moves the peak away from the signal region, which would make a polynomial fit of the background
shape more reliable. However, the gain in the number of background events passing the cuts acts in the opposite direction,
and after studying a $p_T$ cut of 20 GeV, it appears marginal as to whether one would want to lower the $p_T$ cuts to try and
gain sensitivity.
The reconstructed $W$ mass from the signal samples is 
 plotted in Fig.~\ref{fig:sigvar}, for values of $R_{iso}$ used in the below analysis. 

We now turn to discussing the observation prospects of these decays at
the HL-LHC, with 3ab$^{-1}$ of data.  Of order $10^9$ $\ttb \to
W^\pm l^\mp\nu_l b \bar{b}$ events passing the $\ttb$ selection cuts are expected,
where $l$ is an electron or a muon. The number of $\ttb \to W^\pm l^\mp\nu_l b
\bar{b}$ events must be multiplied by the branching ratio of $W\to$
hadrons and $W\to\tau\nu_\tau$ to obtain the number of $W$-had and $W$-tau events, 
and by the branching ratio of $W\to \pi\gamma$ and $W\to \pi\pi\pi$ to obtain the
number of signal events in each decay mode. We use the approximation
that the analysis for the negatively charged $W^-$ signal decay simply
gives a factor two in statistics and, given the difficultly in obtaining events 
at larger values of $R_{iso}$, we assume the $R=0.06$
background shape for both backgrounds in the $\pi
\gamma$ analysis (we find their shapes remain reasonably 
stable as $R_{\text{iso}}$ is increased -- see Fig.~\ref{fig:main}).  For $W\to \pi\gamma$, given that the number of $W$s
passing the $\ttb$ acceptance cuts is of order $\sim10^{9}$, and
calculations of the standard model branching ratio are in the region
$\sim10^{-9} - 10^{-6}$, clearly the HL-LHC could only possibly have
sensitivity to this decay in the upper region of this window. Given
the spread in theoretical predictions, and to give an idea of the
level of exclusion limit the HL-LHC could set, we estimate the value
of branching ratio which would allow for a $3\sigma$ signal,
estimating the significance with
$N_{\text{sig}}/\sqrt{N_{\text{bkg}}}$, where $N_{\text{sig}}$,
$N_{\text{bkg}}$ are the number of signal and background events in the
region $78-84\,$GeV of the $M_{\pi\gamma}$ distribution. 
Optimising over $R_{iso}$ we find that for a value of $R_{iso}=0.15$, a 3$\sigma$ 
discovery is obtained for a
branching ratio of $6\times10^{-7}$, where $N_{\text{sig}}=70$ and
$N_{\text{bkg}}\simeq450$. The number of events in the tail above the
region $78-84\,$GeV is $\sim3000$, meaning that the statistical 
error considered above dominates. This assumes no displaced vertex tagging
applied to the $W$-tau background but, given the one-particle decay mode of the tau, this
is presumably reasonable.
 Even though this branching ratio is 
still above the likely standard model prediction, exclusion limits could be
set lower than the current best limits set by CDF.  For $W\to
\pi\pi\pi$ the entire suppression of background for $R_{iso}\ge0.06$
found in the three pion decay channel means that a discovery estimate
can be translated directly: a sensitivity to a branching ratio of a
few$\times10^{-7}$, which probes well inside the expected standard
model region.

For this analysis we have used leading order event generation, and made no estimate of the theoretical
uncertainties in doing so, although as the analysis is shape driven we do not expect NLO QCD effects to have
a large effect on this.
We also do not take into account any of the realistic collider effects, in particular
the problem of pileup and detector effects. A full study of these effects is beyond the scope of this note, but 
we point out some important handles and improvements which can be made in the analyses which we hope will
ameliorate the inevitable degradation of the results presented here. Firstly, important information is contained in the
direction of the three-momentum of the particles -- particularly well measured for the charged pions -- since these tracks should
point back to the interaction vertex (flagged by the lepton in the $\ttb$ event) coming as they do directly from the $W$ decay. 
This can be used to kill background coming from secondary isolated pion production (such as tau decay) and to 
help deal with pileup contamination -- the vertex must be the same as that determined for the $\ttb$ event. 
Whether this alone will be enough to control pileup and whether new experimental techniques can be
invented to increase sensitivity to exclusive hadronic decays under LHC pileup conditions remain questions
to be answered by a detailed study.
Secondly, given
that the background determination will be data driven, a useful observation is that 
the sign of the lepton coming from the tagging side of the $\ttb$
event fixes the sum of the charges of the signal decay products. Events passing the single particle cuts with the wrong charge sum
provide important information on the nature of the background, even in the signal region. Thirdly, the mass of 
the single particle jets and one of the $b$-jets can be required to be in the vicinity of the top mass, which will further
suppress  background.
Finally, electromagnetic calorimeter profiling can be
used to tighten the definition of `single $\gamma$'-jets above by discriminating between true single, hard photons and 
multiple photons (for example, as is done in $h\to\gamma \gamma$ analyses to suppress $\pi^0\to\gamma \gamma$ contamination).

The above analysis carries over directly to two and three particle decays where charged pions are replaced by charged kaons or
(anti-)protons, since these too are stable as far as the detector is concerned. In reality,  particle identification (PI) is done 
on a statistical basis, so these measurements would overlap into each other.
Similar to the wrong charge sum exploration of the background, groups of decay products 
are forbidden, for example $W^+\to p \pi^+ \pi^-$, although this
has to be convoluted with the uncertainty in PI described above. 
Three particle $W$ decay to charmed mesons, e.g. $W^+\to \{D^+,D^+_s,\ldots\} \pi^+\pi^-$,
and a whole spectrum of higher spin mesons and
baryons could be envisaged, but unlike the pions, kaons and protons,
these particles decay before reaching the detector.  This could give
rise to distinctive signatures, and it would be interesting to
investigate them, in particular the details of the
complications arising from neutrinos in the decays, and the way in
which tagging techniques would fit with the isolation techniques used
here. For example, jet substructure techniques can look for particular
decay patterns inside the cone of size $R_{iso}$ and not make the
single particle veto if a match is found. However, as the results 
 here indicate that observation in the simpler `stable' hadron modes
 will be challenging, we think it unlikely that 
 such measurements would be feasible given the additional experimental difficulties
 they entail.

We briefly comment on the possibilities of precision mass measurement
of the $W$ boson in the three pion decay channel, since a fully
visible final state makes it possible to directly construct a mass
peak, in contrast to the usual techniques that have to deal with
missing energy originating from a neutrino in a leptonic decay
channel.  The current uncertainty on the $W$ mass is $15\,$MeV~\cite{Aaltonen:2013iut}, and the most
precise determination is obtained by fitting the transverse mass
distribution in the lepton-neutrino decay channel at the Tevatron. 
The mass resolution here should be very good, with the
average $p_T$ of the hardest positively charged pion $\langle
p_T\rangle\sim 60\,$GeV, and the $p_T$ of the two other pions 
sharply peaked towards the cutoff $p^{jet}_{T~min}=25\,$GeV, 
 with an average  $\langle
p_T\rangle\sim 40\,$GeV. However, the statistical uncertainty scales like
$\Delta M \sim \Gamma_W/\sqrt{N_{sig}}$, where $\Gamma_W=2.085\,$GeV
is the width of the $W$, and so, to obtain a competitive level of
precision, of order $10^4$ events are necessary, beyond the reach
of the HL-LHC, even with the most optimistic branching ratio of
$10^{-5}$. It would be possible to investigate the use of higher
multiplicity exclusive particle final states, which have substantially
larger branching ratios, but this increases the chances of these
particles falling out of the detector geometry, or falling into the
$b$-jets and other QCD jets.

\section{Conclusions and outlook}
\label{sec:disc}

Top quark pair production provides a potential high-statistics environment
for studying the properties of $W$ boson decays with a limited trigger
bias. Triggering on two $b$-jets and the leptonic decay of one $W$
suppresses QCD backgrounds to both the trigger and the analyses, 
and requiring (transverse) top mass
reconstruction, QCD backgrounds can be even further reduced.  Around
$10^9$ additional $W$s on the other side of the event will be produced
in this way after the HL-LHC run. In this note we have discussed
making measurements of exclusive hadronic decays of the $W$ bosons in
this environment.  We showed that, by using isolation cuts embodied by 
single particle jets, it is possible that the LHC reaches the sensitivity required for
 measuring what would be the first exclusive hadronic decay of a fundamental standard
model boson. We considered as an explicit example the decays $W^+\to
\pi^+\gamma$ and $W^+\to \pi^+\pi^+\pi^-$, and concluded that while the
two-particle decay has a branching ratio which is likely too small for
observation, the three-particle decay has potential to be
measured after the HL-LHC run.

However, a detailed and realistic experimental simulation is necessary to
 determine whether the conclusions presented in this note are
 robust. We have pointed out further experimental handles that can be
 used to improve aspects of the analysis.  It will be interesting to
 see whether single particle jet isolation, used here primarily for
 its simplicity, is a useful technique after detector effects and
 pileup are taken into account. We expect that more usual isolation
 criteria (such as requiring hadronic activity of less than some
 energy in a cone around a particle, similar to those already employed
 in the $h \to\gamma\gamma$ analyses and in tau-lepton identification)
 can be equally well employed to search for these exclusive hadronic
 states. It will also be interesting to see if isolation requirements can
 be useful in more hostile environments, and investigate further
 scenarios where a trade-off in high luminosity in favour of isolated
 signals is beneficial in extracting new measurements.

Although semi-exclusive hadronic measurements of the
form $W\to P X$  (where $P$ is a named particle, and $X$
is anything)  are very challenging in generic $W$ production at a hadron collider, these 
decays also have potential to be studied in the $\ttb$ 
environment. A natural extension of this work would be to investigate
the degree to which a $W$ semi-exclusive decay table, akin to the entries in
the  $Z$ decay table, could be built
up during the course of the HL-LHC run. This would again exploit the fact that after
the $\ttb$ tagging procedure a $W$ boson remains in the event, to which one could
assign particles $P=J/\psi, D^{\pm},B_S^0$, etc, if they are subsequently observed. 
In principle, such measurements could be easier than the fully exclusive decays 
studied here due to their considerably larger branching ratios.

Experimental observation of an exclusive hadronic decay mode of the $W$ at
a hadron collider would bolster proposals in the literature to search for exclusive Higgs decays at
these machines.  It is tempting to speculate on the implications of this study for such measurements, 
in particular the preference for three-body decay modes, due to both the increased branching ratio and 
the background reduction seen here. However, the decay mechanism is different enough to warrant
further study in this direction and,  more importantly, the triggering requirements for dealing with the collider
background are very different between what we study here and  the case of Higgs production\footnote{Of course $\ttb H$ production would have
similar triggering requirements, but the cross section is around three orders of magnitude smaller than
$\ttb$ production,
 reducing sensitivity to rare branching ratios.}. We leave
such considerations to future work. It is clear, however, that if experimental techniques were honed
so as to measure an exclusive decay of the $W$, this would be invaluable in assessing future 
exclusive Higgs decay prospects.

The branching ratios for the exclusive hadronic decays considered here are pushing the limits of the statistics available
at the LHC. 
At a future hadron collider, such as a 100 TeV $pp$ collider, up to two orders of magnitude more
$\ttb$ events are expected. The details of the detectors and experimental methods
for dealing with pileup and hugely energetic particles are completely open ended. However, the bulk of the
$\sim10^{12}$ $\ttb$ events will be produced close to threshold, such that the dynamics of the events themselves
will be very similar to LHC events. 
Furthermore, backgrounds which are not $gg$ initiated 
will not grow as fast as the $\ttb$ cross section. It seems justified
to assume that the reach of such a machine can be estimated by scaling
with the additional luminosity
the results obtained here, accessing
the region of the two particle $W\to \pi \gamma$ and related decays. As a very rough estimate, repeating the 
analysis with  $10^{11}$ $W$s, a 3$\sigma$ observation of $W\to\pi\gamma$ with a branching ratio of $6\times10^{-8}$ is found,
and a $W\to\pi\pi\pi$ decay with a branching ratio $\sim10^{-7}$ should yield a few thousand events.
 These numbers could in principle
 compete with the reach of the proposed future circular
 electron-positron colliders, that should collect clean samples of
 $O(10^8)$ W bosons.

A rich amount of possibilities for extending the known $W$ decay table
lies in exclusive hadronic decays alone.  But, akin to the entries in
the $Z$ decay table, this can be bolstered further already at the LHC,
through searches in the $\ttb$ environment for lepton flavour and
number violating $W$ decays\footnote{For example, see e.g. \cite{BarShalom:2006bv}
  for a study of the parameter space of heavy Majorana neutrinos which
  can be accessed through on-shell $W$ decay.}, and improvements in
the precision of the branching ratios to leptons.  Finally, $W$ boson
properties are just one aspect of the utilisation of the very high
statistics in a $\ttb$ environment.  Because roughly a ninth of $W$s
decay into taus, and another third decay into charmed hadrons, a
similar number of these particles as $W$s opens up the possibility of
a detailed study of their properties in turn. Furthermore, as
discussed in Ref.~\cite{Gedalia:2012sx}, the $b$ quarks produced in
the top decay create an enormous number of $B$-hadrons, which can have
their $b$ or $\bar{b}$ nature determined via the sign of the lepton
from the decay of the associated $W$, after (transverse) top mass
reconstruction.  Looking to the far future, there is a very open
playing field as to the details of new hadron colliders, with plenty
of room for innovative searches and detectors.  In this context, we
look forward to further work into addressing the question: can huge
statistics in the bush compete with a smaller number of clean
events in the hand?

\section*{Acknowledgements}
This work was performed under the framework of  ERC grant  291377 ``LHCtheory: Theoretical predictions and
analyses of LHC physics: advancing the precision frontier''.

\appendix
\section{OPE for the  decay rate $W\to \pi\gamma$}

Here we review Manohar's calculation of $Z\to W\pi$, \cite{Manohar:1990hu}, simply making
 the replacement of $Z$ with $\gamma$ (and using crossing symmetry) to apply it to
  process of interest here, and we compare with the AMP result.
  
\begin{figure}
\begin{center}
\includegraphics[width=9.5cm]{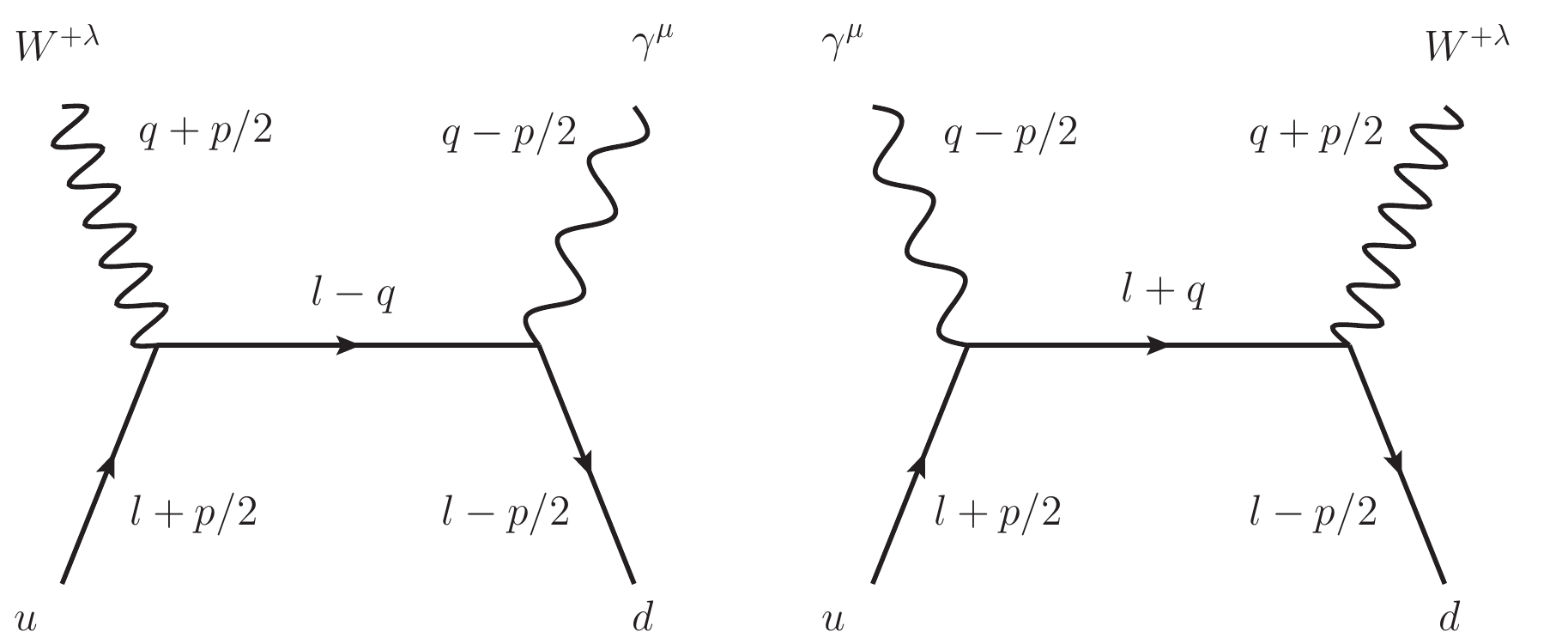}
\end{center}
\caption{Leading order in QCD diagrams for the OPE calculation of the contribution to the decay $W\to \pi \gamma$
shown in Fig.~\ref{fig:amps}~(b).}
\label{fig:ope}
\end{figure}

The contribution to
Fig.~\ref{fig:amps}~(b) at leading order in QCD  involves a calculation of
\be
\int{d^4 x \,e^{i (q-p/2)\cdot x}\, \langle \pi(p) | T[ J_W^\lambda(0) J_\gamma^\mu(x)] | 0 \rangle} = ~~~~~~~~~~~~&~&\nonumber \\
i\bar{d}\left(  Q_d \gamma^\mu \frac{\slashed{l}-\slashed{q}}{(l-q)^2} \gamma^\lambda + Q_u \gamma^\lambda \frac{\slashed{l}+\slashed{q}}{(l+q)^2} \gamma^\mu \right)P_L u &~&\nonumber \\
\ee
using the kinematics shown in Fig.~\ref{fig:ope}. This is  expanded in a power series 
in the variable $\omega = -2 p\cdot q/q^2$, and only the leading twist two operators are retained, defined as
\be
O_L^{\mu_1\ldots\mu_n} = \text{Sym}\left[(i/2)^{n-1} \bar{d} \gamma^{\mu_1} \overset\leftrightarrow{D}\,^{\mu_2}\ldots \overset\leftrightarrow{D}\,^{\mu_n}P_L u\right] \nonumber \\
\ee
where $\text{Sym}[. .]$ symmetries the expression with respect to the Lorentz indices. Defining the matrix elements
of these operators between the pion and vacuum as
\be
\langle \pi (p) | O_L^{\mu_1\dots\mu_n} | 0 \rangle \equiv a_n\,f_\pi p^{\mu_1}\ldots p^{\mu_n} 
\ee
we have $a_1=1/\sqrt{2}$ and all $a_n=0$ for $n$ even\footnote{The first relation follows from the definition of
the pion decay constant, and the second through invariance under CP and a $180^\circ$ rotation in isospin space.}.
For a real  photon of momentum $k=q-p/2$ and a real $W$ of momentum $p_W=p+k=q+p/2$, the polarisation vectors satisfy
\be
\ep\,(k) \cdot k &=& 0 \,,\\ 
\ep\,(p_W) \cdot p_W &=&0 \,.
\ee
This OPE determines the contribution Fig.~\ref{fig:amps}~(b) to be
\be
&~&A_{\text{3(b)}}= V_{ud} \frac{i g}{\sqrt{2}}(-i e)\ep_{\mu}(k)\ep_{\lambda}(p+k) f_\pi \left(   \frac{ a_1g^{\mu\lambda}\,p\cdot q}{q^2}   \right. \nonumber \\
&~& \left.+( p\cdot k\,g^{\mu\lambda}  - p^\mu k^\lambda ) \,\frac{1}{q^2}\,\sum_{n=2}^{\infty}  [Q_u+(-1)^nQ_d] a_n \omega^{n-1} \right. \nonumber \\
&~&\left.- i \epsilon^{\mu\lambda\rho\sigma} p^\rho k^\sigma  \,\frac{1}{q^2}\,\sum_{n=1}^{\infty}  [Q_u-(-1)^nQ_d] a_n \omega^{n-1}  
\right) \,.
\label{eq:a5}
\ee

Returning to the contribution
from Fig.~\ref{fig:amps}~(a), $A_{\text{3(a)}}$, we have,
\be
 A_{\text{3(a)}} = V_{ud} \frac{i g}{\sqrt{2}}(-i e)\ep_{\mu}(k)\ep_{\lambda}(p+k) \frac{f_\pi}{\sqrt{2}}  \left(  \frac{2 g^{\mu\lambda} \,p\cdot q }{p^2-m_W^2 } \right)\nonumber \\
\label{eq:a4}
\ee
 The first term in eq.~\ref{eq:a5} cancels with eq.~\ref{eq:a4} in the full amplitude,
and we find agreement with the general Lorentz structure as in AMP,
\be
&~&A_{\text{3(a)}}+A_{\text{3(b)}} = \frac{-e g V_{ud}}{\sqrt{2}} \ep_\mu(k)\ep_\lambda(p+k)\times \nonumber \\
&~&\left[ \mathcal{A}_\pi (m_W^2) (p\cdot k\, g^{\mu\lambda} - p^\mu k^\lambda ) + \mathcal{V}_\pi(m_W^2) i  \epsilon^{\mu\lambda\rho\sigma} p_\rho k_\sigma  \right] \nonumber \\
\ee
with our expressions for the axial- and vector-like form factors $ \mathcal{A}_\pi$ and $ \mathcal{V}_\pi$ can be read off from eq.~\ref{eq:a5}. The expansion parameter 
\be
\omega=\frac{ 2(m_\gamma^2-m_W^2)}{(m_\gamma^2+m_W^2)}=-2,
\ee
and so using the first term in this
series can only be seen as giving an order of magnitude estimate for the decay rate -- higher order
corrections are clearly important. The situation is analogous to that of the calculation
of $Z\to\pi\gamma$ in \cite{Manohar:1990hu}, for which the expansion parameter takes the same value.
Taking just the leading term,
 the spin averaged
rate obtained is
\be
\Gamma(W\to \pi \gamma) = \frac{\pi \alpha^2 |V_{ud}|^2 f_\pi^2}{54 m_W \sin \theta_W^2} \sim 10^{-9}\, \text{GeV}
\ee
This is smaller than the AMP result by a factor of 2/9. AMP argued $ |\mathcal{A}_\pi(s)/ \mathcal{V}_\pi(s) |  \underset{s\to\infty}{\to} 1$,
taking the form of the vector-like form-factor from the BL asymptotic limit $ \mathcal{V}_\pi(s)   \underset{s\to\infty}{\to} - \sqrt{2}f_\pi/s$.
Since the decay rate is proportional to $|  \mathcal{V}_\pi|^2 (1+ |\mathcal{A}_\pi/ \mathcal{V}_\pi |^2 )$ a factor of 1/2 arises
here, since in the above we have in the leading term $\mathcal{A}_\pi=0$.
A further factor of 2/3 in the amplitude provides a 
further factor of 4/9 in the decay rate. A factor of 2/3 difference with the BL formula (as used by AMP) 
that is  found by taking the leading term in the approach above
was  noted by Manohar in  \cite{Manohar:1990hu}; taking this into account, the two approaches give consistent results. 

\bibliography{rareW.bib}

\end{document}